\documentclass[apj,twocolumn]{emulateapj}

\usepackage{amsmath}

\slugcomment{Accepted to ApJ}

\shortauthors{Parmentier}
\shorttitle{The Multiple Star Formation Efficiencies per FreeFall Time}

\newcommand\effint{\epsilon_{\rm ff, int}}
\newcommand\effmeas{\epsilon_{\rm ff, meas}}
\newcommand\effmeasdt{\epsilon_{\rm ff, meas, \Delta t}}
\newcommand\effcor{\epsilon_{\rm ff, cor}}
\newcommand\effintTH{\epsilon_{\rm ff, int}^{TH}}

\newcommand\tff{\tau_{\rm ff}}

\newcommand\Ms{M_{\odot}}

\newcommand\Msppp{M_{\odot} \cdot pc^{-3}}

\newcommand\cc{cm^{-3}}

\newcommand\fft{freefall time }
\newcommand\sfe{star formation efficiency }
\newcommand\sfr{star formation rate }
\newcommand\stf{star formation }
\newcommand\sfing {star-forming }

\begin{document}


\title{Molecular clumps disguising their star-formation efficiency per free-fall time: What we can do not to be fooled}


\author{Genevi\`eve Parmentier\altaffilmark{1}}


\altaffiltext{1}{Astronomisches Rechen-Institut, Zentrum f\"ur Astronomie der Universit\"at Heidelberg, M\"onchhofstr. 12-14, D-69120 Heidelberg, Germany}


\begin{abstract}
The presence of a volume density gradient in molecular clumps allow them to raise their \sfr compared to what they would experience were their gas uniform in density.  This higher \sfr yields in turn a higher value for the \sfe per freefall time that we measure.  The measured \sfe per \fft $\effmeas$ of clumps is therefore plagued by a degeneracy, as two factors contribute to it: one is the density gradient of the clump gas, the other is the intrinsic \sfe per \fft $\effint$ with which the clump would form stars should there be no gas density gradient.  This paper presents a method allowing one to recover the intrinsic efficiency of a centrally-concentrated clump.  
It hinges on the relation between the surface densities in stars and gas measured locally from clump center to clump edge.  Knowledge of the initial density profile of the clump gas is not required.
A step-by-step description of the method is provided as a tool in hand for observers.      
Once $\effint$ has been estimated, it can be compared with its measured, clump-averaged, counterpart $\effmeas$ to quantify the impact that the initial gas density profile of a clump has had on its \stf history.  

\end{abstract}



\section{Introduction} \label{sec:intro}
Molecular clumps whose gaseous component presents a volume density gradient experience a higher star formation rate than if their gas was of uniform density \citep{tan06,gir11a,elm11,par14p,par19}.  This property stems from the inner-regions of centrally-concentrated clumps forming stars at a pace faster than expected based on the clump mean freefall time.  Their density is actually higher than the clump mean density.  \citet{par19} introduces the notion of {\it magnification factor}, $\zeta$, defined as the factor by which the density gradient of a clump inflates its \stf rate.  In other words, this is the ratio between the \sfr of a clump with a density gradient, $SFR_{clump}$, and the \sfr that the same clump would present in the absence of it, $SFR_{TH}$ ('TH' stands for the 'top-hat' profile of uniform-density gas).  She shows that $\zeta$ depends on the power-law slope of the gas density profile and on the fractional extent of its central core: the steeper the slope, the smaller the radius of the central core as compared to the clump radius, the higher the magnification factor $\zeta$.

The density gradient inside molecular clumps also impacts their \sfe per \fft as we measure it, since it depends on the clump \stf rate $SFR_{clump}$ as \citep[see e.g.][]{kt07,eva09,lad10,mur11,kru12,vut16,och17}:
\begin{equation}
\effmeas = \frac{SFR_{clump} \langle \tff \rangle}{m_{gas}}\;.
\label{eq:effmeas}
\end{equation}
In this equation, $m_{gas}$ and $\langle \tff \rangle$ are the gas mass and gas mean free-fall time of the clump, respectively.  A density gradient therefore also inflates the \sfe per \fft as compared to what would be measured for a top-hat profile.  
\citet{par19} refers to the efficiency that would be measured for a top-hat profile as the {\it intrinsic} \sfe per free-fall time, $\effint$, to emphasize that $\effint$ is unaffected by the density gradient.  $\effint$ also characterizes the \stf activity of the nested shells of gas of which the clump is made, as long as these shells are thin enough to be considered of uniform density \citep[see Equation 4 in][]{par19}.  The measured \sfe per \fft derived from Equation~\ref{eq:effmeas} by observers therefore constitutes a global quantity, while its intrinsic counterpart is a local quantity.  How to reveal  the latter is the main objective of this paper.     
All these parameters are intertwinned through her Equation~10, which we reproduce here for the sake of clarity:
\begin{equation}
\zeta = \frac{SFR_{clump}}{SFR_{TH}}=\frac{\effmeas}{\effint}\;.
\label{eq:zetadef}
\end{equation}
A crucial consequence is that even if the intrinsic \sfe per \fft $\effint$ were universal (that is, no variations from clump center to clump edge, and no inter-clump variations), the measured \sfe per \fft $\effmeas$ would present wide fluctuations.  Such fluctuations, embodied by the $\zeta$ factor, reflect the diversity in clump structures rather than variations in the physics of star formation.  
Observations of molecular clumps of the Galactic disk show that the logarithmic slope of their power-law density profile varies from $\simeq-1$ down to $\simeq -4$ \citep[e.g.][]{mue02,schn15}.  The steepest density profiles (slope equal to or steeper than $-3$) are observed in dense pc-size regions of nearby molecular clouds \citep[][]{schn15}.  For such steep density profiles, \citet{par19} shows that the magnification factor can boast more than one order of magnitude depending on the central peakedness of the gas density profile, that is, depending on the profile slope and on the extent of the central core (see her Section~5).    

Global observations of molecular clumps therefore leave us with an annoying degeneracy, for how can we disentangle in the measured \sfe per \fft $\effmeas$ the impact of the density gradient (i.e. $\zeta$) from the contribution of the \stf process {\it per se} (i.e. $\effint$)? For instance, consider a clump whose measured \sfe per \fft is $\effmeas=0.10$.  How should it be interpreted?  Does it imply that any small region of the clump forms stars with an (intrinsic) \sfe per \fft of $\effint=0.10$?  Or is this efficiency smaller (e.g. $\effint=0.02$), with the "missing" factor $5$ being contributed by the clump density gradient?  

\citet{par19} mapped the magnification factor $\zeta$ of model clumps as a function of time and for a range of conditions at \stf onset.  These conditions are, for each model clump, its mass, radius, volume density profile, and intrinsic \sfe per free-fall time.  In this contribution, we consider the reversed problem: given a model clump at time $t$ after \stf onset, how can we recover its intrinsic \sfe per free-fall time?  That is, the goal is now to quantify $\effint$ in  clumps with ongoing star formation, where the initial gas distribution has been altered by star formation.  We will show how spatially-resolved observations can help us break the degeneracy existing between $\zeta$ and $\effint$, with the additional advantage of using gas and star projected/surface densities, which are more easily measured than spatial/volume densities.  Once estimated, $\effint$ can be combined with the "traditional" measured \sfe per \fft to quantify the impact of the clump density gradient as $\zeta=\effmeas/\effint$.

The outline of the paper is as follows.  In Section~\ref{sec:modsum}, we summarize the key aspects of the model of  \citet{par19}.  Section~\ref{sec:effmeas} shows how the measured \sfe per \fft is obtained for the model clumps.  In Section~\ref{sec:spres}, we explain how spatially-resolved data can help us probe $\effint$, and we introduce our method.  In Section~\ref{sec:test}, we apply it to the comprehensive grid of model clumps built by \citet{par19}.  A step-by-step description of how to implement the method is presented in Section~\ref{sec:recip}.  Section~\ref{sec:disc} contains a brief discussion, followed by the conclusions in Section~\ref{sec:conc}.

\section{Model-predicted magnification factor}\label{sec:modsum}
In this section, we summarize the model implemented by \citet[][hereafter Paper~I]{par19} and how it predicts the magnification factor of model clumps.    

The volume density profile of a model clump is described by a decreasing power-law of logarithmic slope $-p_0$ with a central core $r_c$ so as to avoid a density singularity at the clump center (see Equation~11 in Paper I).  The clump mass $m_{clump}$ is the mass enclosed within the radius $r_{clump}$.  The clump density profile is also the density profile of the clump gas at \stf onset ($t=0$).  Equation 19 in \citet{par13} allows one to obtain the corresponding gas density profile at time $t$ after \stf onset for a given intrinsic \sfe per free-fall time.  The latter is assumed to be time-invariant and $\effint = 0.01$ is assumed in all simulations.  The \sfr $SFR_{clump}$ of the clump is then obtained by numerically integrating the \sfr of nested shells of gas from clump center to clump edge, the \stf activity of all shells being characterized by the intrinsic \sfe per \fft (see Equation~4 in Paper I).  The gas mass $m_{gas}$ of the clump at any time $t$ is similarly obtained by integrating the corresponding gas volume density profile over the clump volume.  The difference between the total mass of the clump $m_{clump}$ (equivalently the gas initial mass $m_{gas}(t=0)$) and the gas mass at time $t$ provides the mass in stars formed over the time-span $t$.  The global \sfe (i.e. the fraction of the initial gas mass turned into stars) follows as $SFE_{gl}(t) = m_{stars}(t)/m_{clump}=(m_{clump}-m_{gas}(t))/m_{clump}$.   

It is now doable to predict the corresponding magnification factor $\zeta(t)$.  The model assumes that the clump radius remains constant through the \stf process, and  knowledge of the gas mass $m_{gas}$ thus yields the mean density of the gas $\langle \rho_{gas} \rangle$ and its mean free-fall time
\begin{equation}
\langle \tff \rangle = \sqrt{\frac{3\pi}{32G \langle \rho_{gas} \rangle}}\;.
\end{equation}        
The \sfr that the clump {\it would} experience in the absence of a density gradient is then given by: 
\begin{equation}
SFR_{TH} = \effint \frac{m_{gas}}{\langle \tff \rangle}\;.
\end{equation}      
It stems from assuming a constant gas density in Equation~4 of Paper~I.  
The model-predicted magnification factor then follows from its definition, namely, $\zeta = SFR_{clump}/SFR_{TH}$.  

Color-coded maps of the magnification factor at $t=0$ and $t=0.5$\,Myr are presented in Figures~5, 6 and 7 in Paper~I for various clump masses, radii, and initial volume density profiles of the gas.  A more centrally-peaked density profile - be it through a steeper slope or a smaller central core as compared to the clump radius - yields higher \stf rates and higher magnification factors.  Because \stf operates fastest in the high-density regions of the clump center, the gas density profile loses part of its central peakedness as time goes by and $\zeta$ decreases as a result.  The decrease is faster for models with a higher mean volume density (hence a shorter mean freefall time) and/or with a steeper slope initially (thus a higher central density).  
 
To derive the magnification factor of model clumps at any time $t$ is doable because we can predict the gas density profile at that time $t$ based on our assumptions of an initial gas density profile and intrinsic \sfe per free-fall time.  But what about star-forming clumps of the Galactic disk, i.e. clumps observed at time $t>0$ after star formation onset?  Neither their intrinsic \sfe per free-fall time, nor the initial spatial distribution of their gas is known.  If $\effmeas$ is, for instance, high, one cannot say whether this stems from a high $\effint$ or from an initially steep profile (see Equation~\ref{eq:zetadef}).  In addition, the initial gas density profile has been modified, especially in the clump inner regions, precisely the regions boosting early star formation.  Therefore, the current gas density profile of a \sfing clump cannot fully shed light on its past \stf history.    

To make progress, we will introduce in Section \ref{sec:spres} a method to estimate the intrinsic \sfe per free-fall time of a star-forming clump, which does not require knowledge of its initial gas density profile.  We will apply it to the model clumps calculated in Paper~I, and we will show that it yields estimates of $\effint$ which are in good agreement with the value actually used in the simulations (i.e. $\effint=0.01$).  Once $\effint$ has been estimated, the impact of the initial gas density profile can be recovered as the ratio $\zeta = \effmeas/\effint$ (Equation~\ref{eq:zetadef}). 
We now describe how we obtain the measured (global) \sfe per \fft of our model clumps. 

\section{Time-averaged measured star formation efficiency per freefall time}\label{sec:effmeas}
The measured \sfe per \fft depends on the clump gas mass $m_{gas}$ and radius $r_{clump}$ (whose combination yields the gas mean density $\langle \rho_{gas} \rangle$ and \fft $\langle \tff \rangle$), and on the clump \stf rate $SFR_{clump}$ (Equation~\ref{eq:effmeas}).  In Paper~I, $SFR_{clump}$ refers to the instantaneous \stf rate, that is, the \sfr at a given time $t$. 
However, for \sfing regions whose young stellar objects (YSO) can be counted, observers often obtain time-averaged \stf rates by combining the total mass in YSOs, $m_{YSO}$, with an assumed duration of the \stf episode $t$, i.e. $\langle SFR_{clump} \rangle = m_{YSO}/t$.  We therefore introduce a more practical definition of the measured \sfe per freefall time    
\begin{equation}
\effmeasdt = \frac{m_{stars}}{t} \times \frac{\langle \tff \rangle}{m_{gas}}\;.
\label{eq:effm2}
\end{equation}  
In the right-hand-side, $m_{stars}$ is the stellar mass built by the model clump within the time-span $t$. 
The subscript $\Delta t$ in the left-hand-side indicates that the measured \sfe per \fft now builds on a time-averaged \stf rate.  Given that the instantaneous \sfr of the model clumps decreases with time (see Figure~4 in Paper~I),  the advantage of using a time-averaged \sfr is that it keeps track of the more vigorous \stf activity experienced at earlier times, thereby better preserving the impact of the initial gas density profile.  This is $\effmeasdt$ that we will  compare to our estimates of the intrinsic \sfe per freefall time (in Section~5).  We now move into how to reveal the latter. 

\section{Spatially-Resolved Observations to the Rescue}\label{sec:spres}
\subsection{Core of the method}   \label{ssec:conc}

Global observations of a clump with a gas density gradient can only yield its {\it measured} \sfe per free-fall time (see Equation~\ref{eq:effmeas}).    
To estimate its intrinsic \sfe per free-fall time, one would ideally have at one 's disposal a second clump containing the same gas mass distributed within the same radius, but according to a top-hat profile.  Its \sfr $SFR_{TH}$ would be measured and its \sfe per \fft would be inferred as:
\begin{equation}
\effint = \frac{SFR_{TH} {\langle \tff \rangle}}{m_{gas}}\;.
\label{eq:effTH}
\end{equation}
However, Nature does not offer us the top-hat equivalent of any \sfing clump we observe, and a method different from that building on Equation~\ref{eq:effmeas} needs to be elaborated.

While the measured \sfe per \fft characterizes the \stf activity of the clump globally, the intrinsic \sfe per \fft that we seek to estimate characterizes the \stf activity of the clump shells, from its edge to its  center (see Equation~4 in Paper~I).  A seemingly naive approach is thus to focus on one of these shells.  We will do just that.  Specifically, we target the shell whose initial gas density is the clump mean density.  We refer to this shell as the "mean shell".  Its density is also the density $\rho_{TH}$ of a top-hat model with identical clump mass $m_{clump}$ and radius $r_{clump}$.  We can write, with $r_m$ the mean shell radius, $\rho_{clump}(r_m)$ the initial gas density of the mean shell, and $\rho_{TH}$ the density of the top-hat model:
\begin{equation}
 \rho_{TH} = \frac{m_{clump}}{\frac{4\pi}{3}r_{clump}^3} = \rho_{clump}(r_m)\;.
 \label{eq:msh}
\end{equation}  

In the model of \citet{par13}, the volume density in stars depends on the initial gas density, on the intrinsic \sfe per \fft $\effint$ and on the \stf time-span $t$ (see their Equations~19 and 20).  That is, the longer the \stf duration $t$, the higher the intrinsic \sfe per \fft $\effint$ and/or the higher the initial gas density (hence the shorter the gas initial freefall time), the faster the gas gets converted into stars and the lower/higher the gas/star density at time $t$.  Given that the mean shell and the top-hat model have the same initial gas density, they evolve at the same pace {\it provided their intrinsic \sfe per free-fall time $\effint$ is the same}. 

\begin{figure}
\begin{center}
\epsscale{1.0}  \plotone{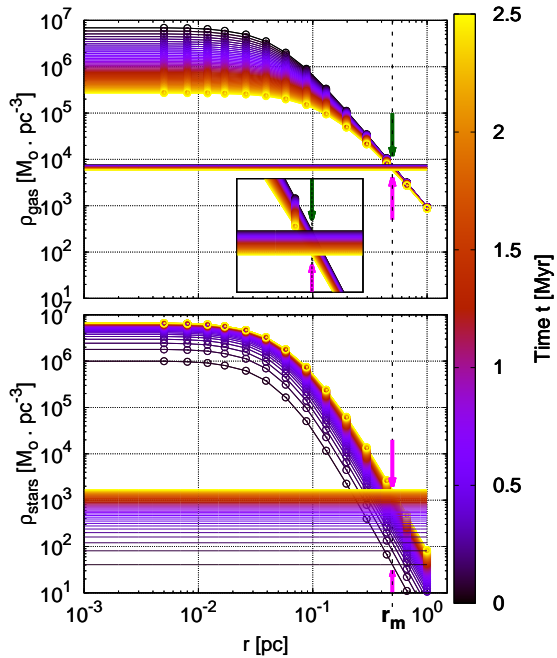}
\caption{Top panel: Evolution with time of the gas density profile of two model clumps, one with a top-hat density profile (symbol-free lines), the other with an initial steepness $p_0=3$ (lines with open circles).  The time $t$ since \stf onset is color-coded by the right-hand-side palette.  Model parameters are: $m_{clump} = 3.2\cdot 10^4\,\Ms$, $r_{clump} = 1$\,pc $\effint=0.01$, and $\rho_c = 7 \cdot 10^6\,\Msppp$.  The region where the local gas densities of both models are initially equal (i.e. the mean shell; green arrows) is zoomed-in.  The vertical dashed line indicates the radial position of the mean shell (radius $r_m$).  The magenta arrow indicates that both densities have remained equal to each other by $t=2.5$\,Myr.  Bottom panel: same for the evolution of the stellar density profiles.  The magenta arrows indicate that the stellar density of the top-hat model and the stellar density of the mean shell of the $p_0=3$ model are equal all through the simulations, from $t=0.05$\,Myr to $t=2.50$\,Myr.}
\label{fig:rhogs}
\end{center} 
\end{figure}

This is illustrated in the top panel of Figure~\ref{fig:rhogs}, which shows the evolution of a gas density profile with $p_0=3$ (i.e. an initially steep density gradient) and of a top-hat model with the same clump mass, radius and intrinsic \sfe per freefall time.  Model parameters are $m_{clump} = 3.2\cdot 10^4\,\Ms$, $r_{clump} = 1$\,pc, and $\effint=0.01$, combined to a gas initial density profile of steepness $p_0=3$ and central density $\rho_c=7\cdot10^6\,\Msppp$.  The clump  mean volume density ($\simeq 8000\Msppp$) therefore falls in the density regime for which steep radial density profiles have been observed in molecular clouds of the Galactic disk \citep[i.e. $10^4\,\cc < n_{H2} < 3\cdot10^5\,\cc \equiv 700\,\Msppp  < \langle \rho_{clump} \rangle < 2.1\cdot10^4\,\Msppp$;][]{schn15}.  
The darkest line with open circles depicts the initial gas density profile $\rho_{clump}(r)=\rho_{gas}(r,t=0)$ and the horizontal darkest line (best visible in the zoom-in region) is the top-hat model.  The vertical dashed line marks the radius $r_m$ of the mean shell, that is, the radius at which the clump density profile equates the top-hat density (as indicated by the downward green arrows).  At any given time $t$ of their evolution, the mean shell and the top-hat profile keep presenting the same gas volume density because of their common intrinsic \sfe per \fft and common initial gas density.  The upward magenta arrows highlight the equality at $t=2.5$\,Myr (best visible in the zoom-in region).  
Similarly, the mean shell and the top-hat profile present the same stellar density at any time $t$ (see bottom panel of Figure~\ref{fig:rhogs} which shows the rise with time of the stellar density profiles of both models).         \\

Knowledge of the radial position $r_m$ of the mean shell is not needed, however.  Only its gas- and star-volume densities are.  In the next section, we therefore move to the ($\rho_{gas}$, $\rho_{stars}$) space.
 
\subsection{From radial density profiles to star formation relations}   \label{ssec:sfl}

\begin{figure}[t]
\begin{center}
\begin{minipage}{1.00\linewidth}
\epsscale{1.0}  \plotone{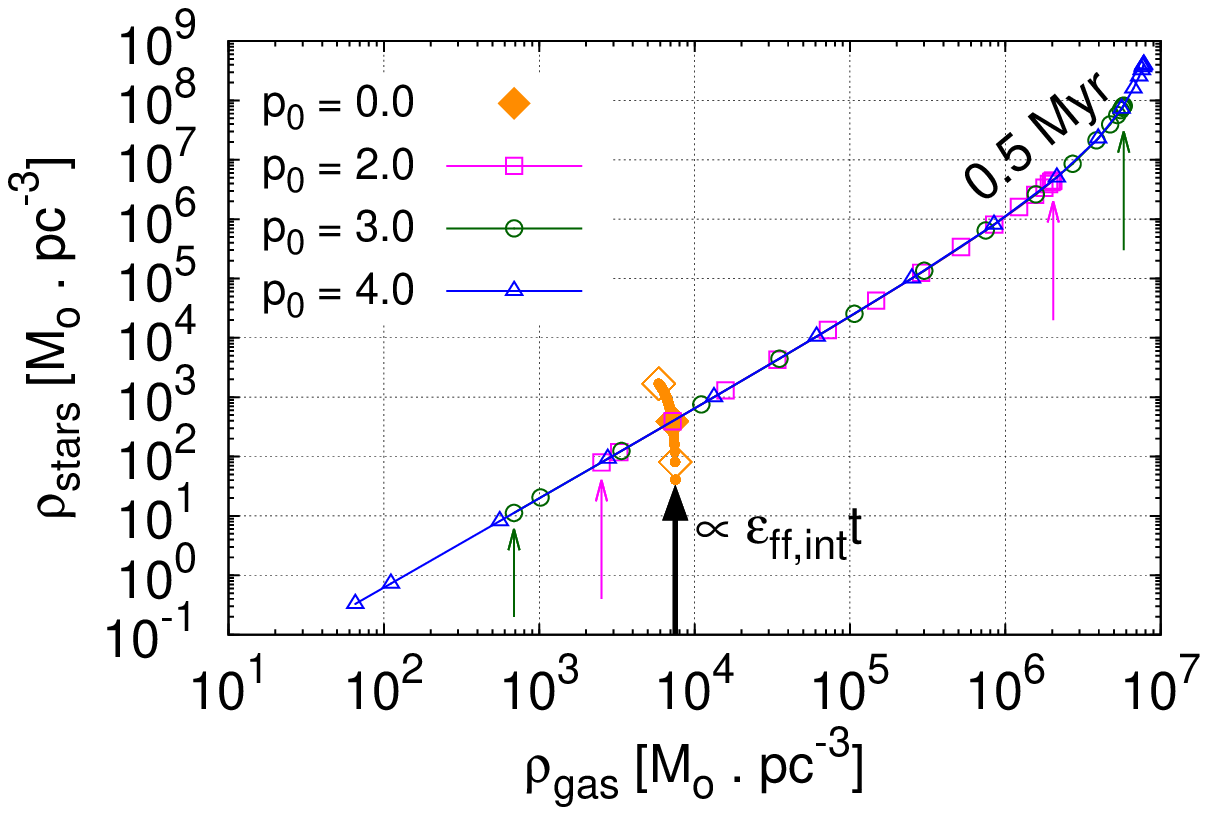}
\end{minipage} 
\vfill
\begin{minipage}{1.00\linewidth}
\epsscale{1.0}  \plotone{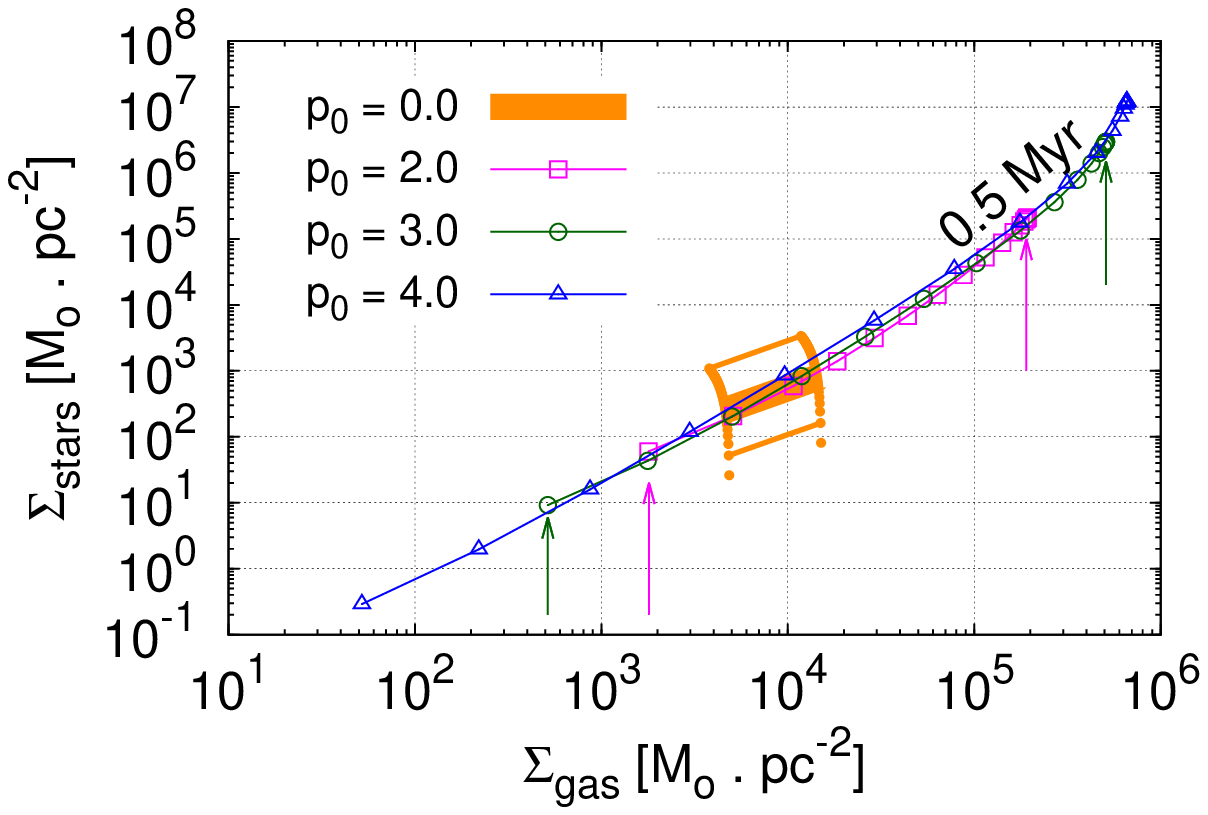}
\end{minipage} 
\end{center}
\caption{
Top panel: Star formation relations of model clumps based on their local volume densities in stars and gas.  Model parameters are $m_{clump} = 3.2 \cdot 10^4\,\Ms$, $r_{clump}=1$\,pc, $r_c=0.02$\,pc, $\effint =0.01$, and $t=0.5$\,Myr, and $p_0$ varying from $0$ (top-hat profile) to $p_0=4$ (very steep profile; see the key).  The extent of the $p_0=(2,3)$ models is shown by the thin vertical arrows, color-coded accordingly.  The orange open diamonds indicate the locus of the top-hat model when $\effint^{TH}=5\effint=0.05$ and $\effint^{TH}=\effint/5=0.002$.  The track of orange plain circles indicate the time-evolution of the model with $\effint^{TH}=0.01$ from $t=0.05$\,Myr up to $t=2.5$\,Myr.   Bottom panel: Same as top panel, but based on the local surface densities in stars and gas.  The orange plain rectangle, the upper and lower orange thick lines indicate, respectively, the locations of the top-hat model at $t=0.5$\,Myr when $\effint^{TH}=0.01, 0.05$ and $0.002$.  
}
\label{fig:sfl1}
\end{figure}

The top panel of Figure~\ref{fig:sfl1} shows the volume-density-based \stf relation for several model clumps, which is the stellar density $\rho_{stars}$ in dependence of the gas density $\rho_{gas}$.  We refer to this relation as a {\it local} \stf relation, since the gas and star densities are those at given distances of the clump center  (i.e. $(\rho_{gas}(r), \rho_{stars}(r))$), rather than densities averaged over the whole clump.  The top-right end of a relation corresponds to the clump central regions, while the lower-left end depicts the clump outskirts.  All four clumps have the same mass ($m_{clump}=3.2\cdot10^4\,\Ms$), radius ($r_{clump}=1$\,pc), \stf time-span ($t=0.5$\,Myr), intrinsic \sfe per \fft ($\effint=0.01$), but four distinct density profiles initially, namely $p_0=0,2,3$ and $4$.  At the clump center, a core radius of $r_c=0.02$\,pc is initially imposed and the gas initial central density $\rho_c$ is adjusted such that the radius $r_{clump}$ contains the mass $m_{clump}$.
As the density gradient steepens, the densities in the clump inner regions increase, while those in the clump outskirts  decrease.  This results in stretching out the relation as $p_0$ increases.  
Each density profile has a mean shell, and all three mean shells have the same initial gas density, which is that of the top-hat model ($p_0=0$).  The top-hat model (orange plain diamond) therefore identifies the location of the mean shell of each centrally-concentrated clump in the ($\rho_{gas}, \rho_{stars}$) space.  That is, it marks the location of a clump region which evolves at the same pace as the top-hat model.  Their evolution is dictated by the {\it intrinsic} \sfe per freefall time, not by the global/{\it measured} one.  

If one runs another top-hat model with the same \stf time-span $t$ but too high (too low) an intrinsic \sfe per free-fall time, then that top-hat model finds itself above (below) the clump local \stf relations.  This is illustrated by the open diamonds for which the intrinsic efficiency is 5 times higher (lower) than used for the $p_0=(2,3,4)$ models.  Therefore, the top panel of Figure~\ref{fig:sfl1} illustrates that if the initial mean density of a clump and the time elapsed since \stf onset are known, its intrinsic \sfe per \fft can be recovered by comparing its local \stf relation to the predictions made for top-hat models of various intrinsic efficiencies $\effintTH$.  Knowledge of the initial gas density profile is not required.   
A limitation of the method as it is now, however, is that it builds on volume densities, while the densities directly measured by observers are surface densities.  Therefore, we now move to the space of projected star and gas local densities $(\Sigma_{gas}, \Sigma_{stars})$.

\subsection{From volume densities to surface densities}   \label{ssec:surf}

Spatially-resolved observations of molecular clouds have revealed their local \stf relation, namely, the relation between the local surface density in young stars and the local surface density of the gas.  As for the model volume densities above, we coin those surface densities "local" because they are measured at the location of individual proto- or pre-main-sequence stars \citep[e.g.][]{gut11}, or measured within given gas-surface-density contours \citep[e.g.][]{hei10}.  They are not averaged over an entire cloud or clump.  With the advent of the Atacama Large Millimeter Array, such diagnostic plots are now also collected for molecular clouds of the Central Molecular Zone \citep[e.g.][]{gin18}.  

\begin{figure*}
\begin{center}
\epsscale{1.0}  \plotone{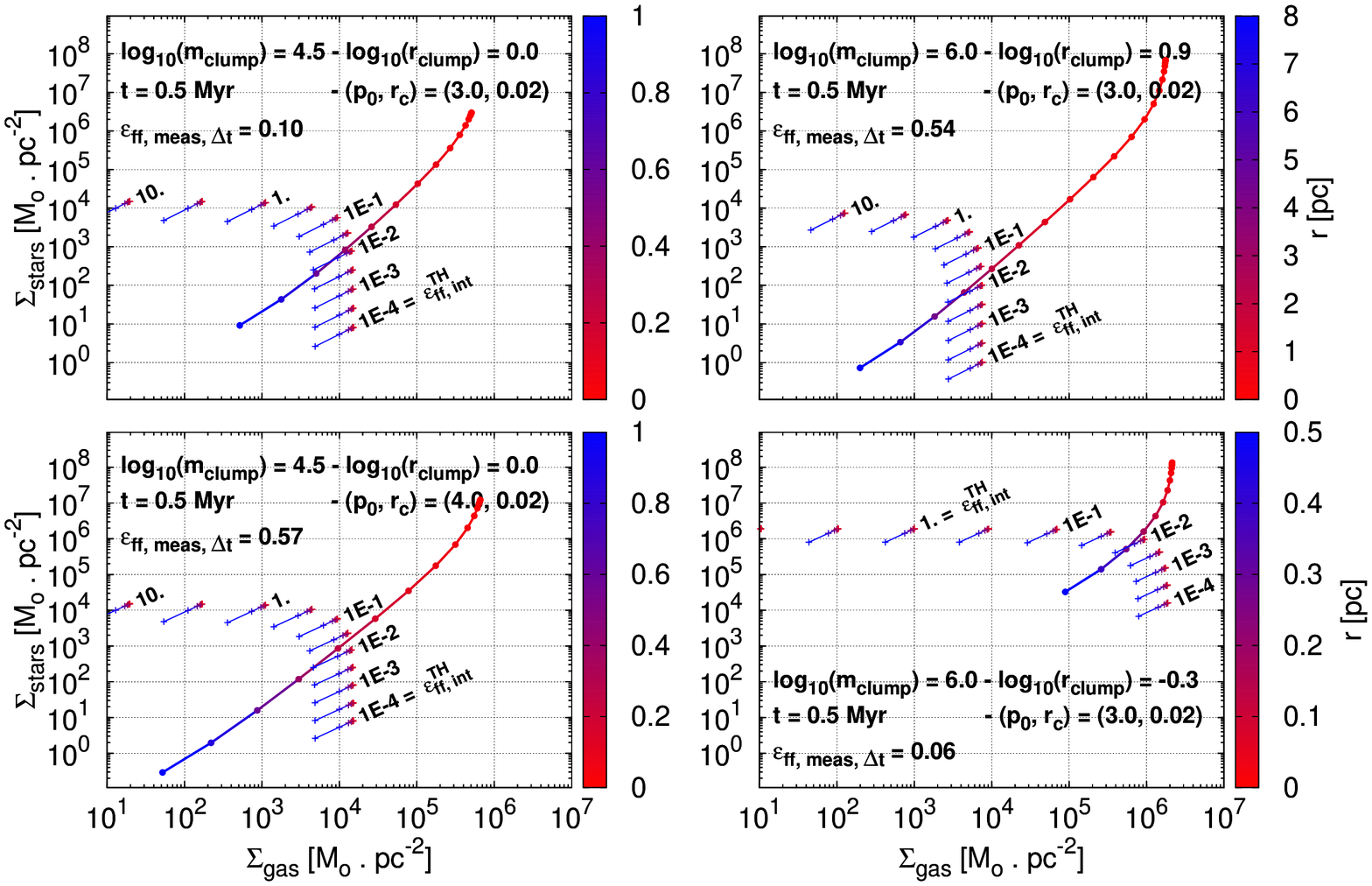}
\caption{
For each panel, comparison between the \stf relation of a centrally-concentrated clump and a grid of \stf relations for top-hat models.  Common model parameters are the radius $r_{clump}$, enclosed mass $m_{clump}$ and \stf time-span $t$.  They are quoted in each panel, along with the parameters $(p, r_c[pc])$ of the gas initial density profile of the centrally-concentrated clump.  Also given is its measured \sfe per \fft $\effmeasdt$ as defined by Equation~\ref{eq:effm2} (i.e. based on a time-averaged \stf rate).  
The \stf relations of the top-hat models have been calculated for intrinsic \stf efficiencies per free-fall time $\effint^{TH}$ ranging from $10^{-4}$ to $10$, in logarithmic steps of 0.5 as quoted next to every two relations.  The comparison of both types of \stf relation yields an estimate of the intrinsic \sfe per \fft of the centrally-concentrated clump ($\effint = 0.01$ in all simulations).   
All relations are color-coded as a function of the distance $r$ to the clump center, each color palette having its own extent as given by the corresponding clump radius.  
  }
\label{fig:corgrid}
\end{center} 
\end{figure*}

The bottom panel of Figure~\ref{fig:sfl1} presents the projected local \stf relations $(\Sigma_{gas},\Sigma_{stars})$ of the models shown in the top panel.  The orange rectangle depicts the top-hat model which, as in the top panel, overlaps with the $p_0 \neq 0$ models as a result of their common \stf time-span, common gas initial mean density, and common intrinsic \sfe per free-fall time.  The thick orange lines above and beneath it depict the models with $\effint^{TH}=5\effint$ and $\effint^{TH}=\effint/5$.

\begin{figure*}
\begin{center}
\epsscale{1.0}  \plotone{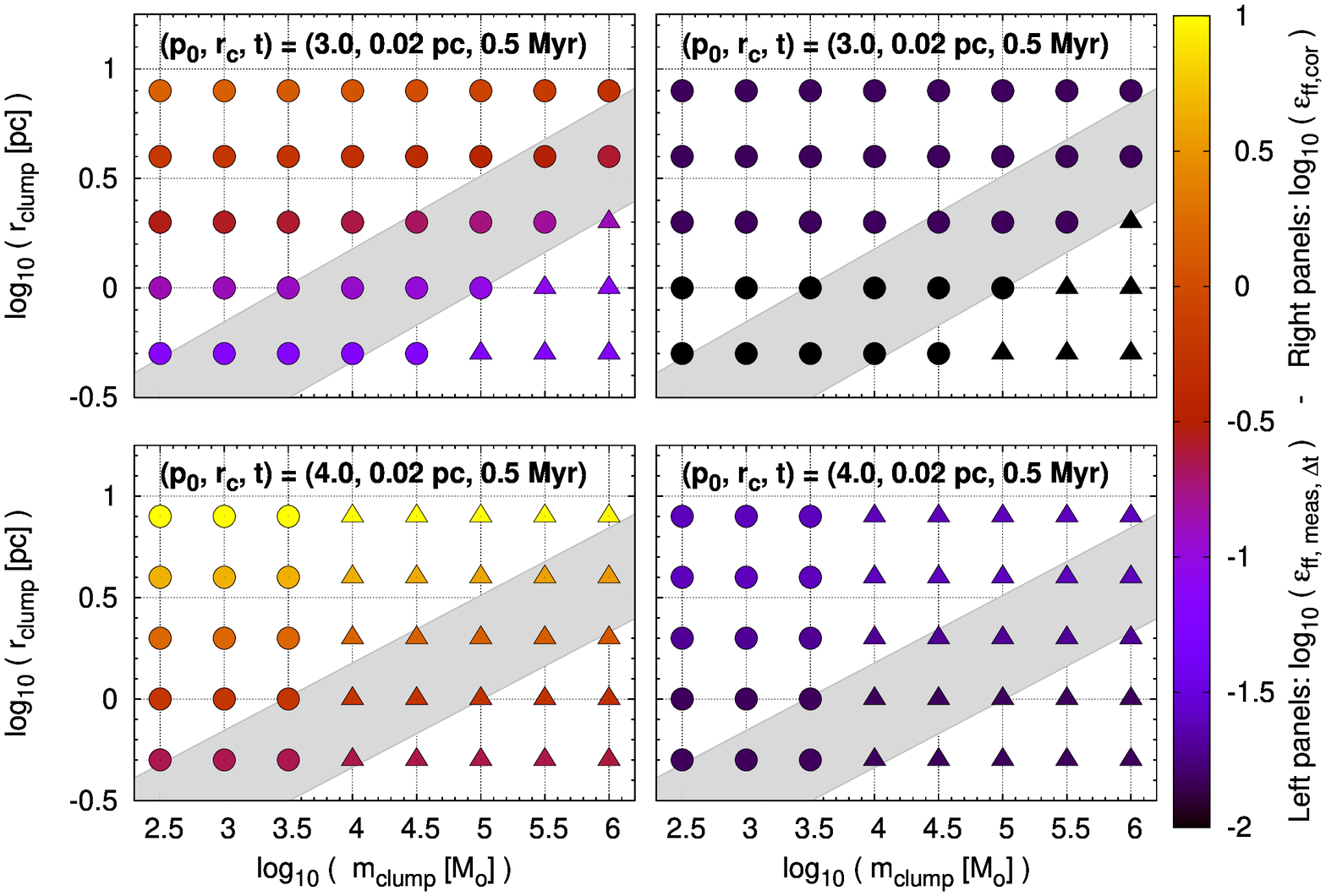}
\caption{
Comparison between the measured \sfe per free-fall time of our model clumps ($\effmeasdt$, left panels) and our estimates of their intrinsic efficiency ($\effcor$, right panels) with the initial core radius imposed.  Parameters are the radius and initial gas mass of clumps, $r_{clump}$ and $m_{clump}$, and the initial steepness $p_0$ of their density profile (top panels: $p_0=3$; bottom panels: $p_0=4$).  
The initial core radius and the intrinsic \sfe per \fft are set to $r_c = 0.02$\,pc and $\effint = 0.01$.  Each model is represented by a plain symbol, the color of which depicts the value of the corresponding efficiency (see palette for color-coding, with a logarithmic scale).  The grey stripe highlights the density regime $10^4\,\cc < n_{H2} < 3\cdot10^5\,\cc \equiv 700\,\Msppp  < \langle \rho_{clump} \rangle < 2.1\cdot10^4\,\Msppp$ for which steep density profiles have been detected in Galactic clouds \citep{schn15}.  Model clumps which by $t=0.5$\,Myr have achieved $SFE_{gl}>0.50$ are depicted by triangles.
}
\label{fig:cormap}
\end{center} 
\end{figure*}

The concept is further illustrated in Figure~\ref{fig:corgrid}.  Each of its four panels compares {\it (i)} the local \stf relation of a clump with a steep density gradient (either $p_0=3$ or $p_0=4$) and an intrinsic \sfe per \fft $\effint=0.01$ to {\it (ii)} a grid of relations obtained for a uniform-density clump ($p_0=0$) with identical initial gas mass, radius, \stf time-span, but different intrinsic \stf efficiencies per free-fall time.  The clump mass and radius, the time-span since \stf onset, and the density gradient of the centrally-concentrated clump are quoted in each panel.  Both types of models ($p_0=0$ and $p_0=(3,4)$) are easily distinguishable based on their distinct extent in the $(\Sigma_{gas}, \Sigma_{stars})$ space.  All relations are color-coded as a function of the distance $r$ from the clump center.  In that respect note that each color palette has its own upper limit, to reflect the size of each model clump.   The intrinsic \sfe per \fft of the top-hat models ranges from $\effint^{TH}=10^{-4}$ to $10$ in logarithmic steps of 0.5 (value quoted to the right of every two relations).  As $\effint^{TH}$ increases (while retaining the same \stf time-span $t$), the corresponding \stf relation moves to higher star- and lower gas-surface densities, highlighting thereby a faster pace of \stf for higher \stf efficiencies per free-fall time.  Also quoted in each panel is the measured \sfe per free-fall time of the centrally-concentrated clump $\effmeasdt$ (see Section \ref{sec:effmeas}).  It sometimes differs from its intrinsic counterpart $\effint=0.01$ by more than an order of magnitude.  Nevertheless, a mere visual inspection of the diagrams yields an estimate of the intrinsic \sfe per free-fall time for the $p_0=(3,4)$ models.  This one is revealed by the top-hat model whose local \stf relation best overlaps the clump model with a steep, but unknown, initial density gradient.  We refer to the $\effint$ estimate as the corrected \sfe per free-fall time, $\effcor$.  The method yields $\effcor \simeq 0.01$ for each of the four cases, in agreement with the intrinsic efficiency actually used in the simulations, i.e. $\effint = 0.01$.  Resolved observations hold therefore the potential to  deliver the right order-of-magnitude for $\effint$, while the measured efficiency provided  by global observations can be off by an order-of-magnitude, or more.      

\section{Application of the method}\label{sec:test}

We have applied the method devised in Section \ref{ssec:surf} to all the model clumps with $p_0=3$ and $p_0=4$ computed in Paper~I.  When $p_0\geq3$, the ratio between the (instantaneous) measured \sfe per \fft and its intrinsic counterpart can reach 3 orders of magnitude (equivalently the magnification factor can reach $\zeta(t) \gtrsim 10^3$; see e.g. bottom-right panel of Figure 5 in Paper~I).  This is therefore the regime where an estimate of $\effint$ is the most needed.  Models with $p_0\geq3$ are of two types (Sections~5.1 and 5.2 of Paper~I, respectively).  In a first category, the central core radius $r_c$ of the gas initial density profile is imposed ($r_c=0.02$\,pc) and the clump central density $\rho_c$ is calculated such that the clump radius $r_{clump}$ contains the assigned clump mass $m_{clump}$.  In a second category, this is the central density $\rho_c$ of the initial gas density profile which is imposed ($\rho_c=7\cdot10^6\,\Msppp$) and the corresponding core radius $r_c$ is inferred such that the clump radius $r_{clump}$ contains the clump mass $m_{clump}$.  In most cases, the first category yields higher magnification factors because the absence of constraint on the central density leads to greater density contrasts between the clump center and the clump edge.  Magnification factors predicted for clumps whose initial gas central density is imposed are in contrast smaller, especially for massive clumps (compare e.g. the bottom-right panel of Figure 5 with the top-right panel of Figure 7 in Paper~I).  This is because a compact massive clump with a limited central density necessarily presents a wide central core to accommodate its large mass $m_{clump}$ inside its given radius $r_{clump}$.  As a result, the impact of the steep density profile in the outer regions is counteracted by the significant flat density profile in the inner regions.  

The model clumps that we consider therefore present a wide range of gas initial density profiles (hence of values of the initial magnification factor $\zeta$), in addition to large ranges of clump radii (from 0.5 to 8\,pc), initial gas masses (from 300 to $10^6\,\Ms$),  and \stf time-spans (up to 2.5\,Myr after \stf onset).  

\begin{figure*}
\begin{center}
\epsscale{1.0}  \plotone{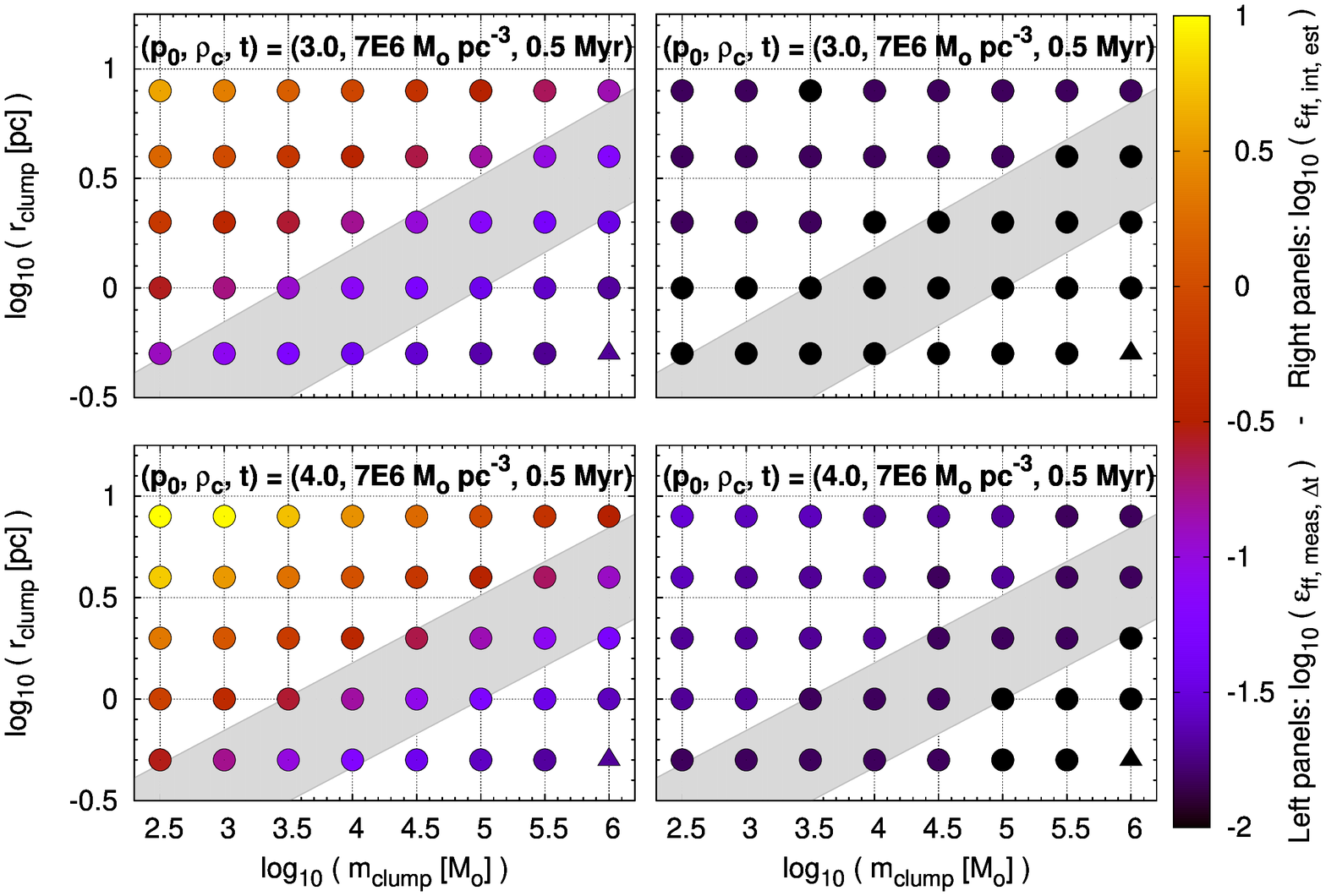}
\caption{Same as Figure~\ref{fig:cormap}, but with the initial central density of the clump gas imposed: $\rho_c = 7\cdot10^6\,\Msppp$ }
\label{fig:cormap2}
\end{center} 
\end{figure*}

The left panels of Figure~\ref{fig:cormap} present the measured \sfe per \fft $\effmeasdt$ (Equation~\ref{eq:effm2}) for model clumps with the initial core radius $r_c$ imposed.  The initial steepness of the gas density profile is either $p_0=3$ (top panel) or $p_0=4$ (bottom panel).  The \stf time-span, the initial core radius and the intrinsic \sfe per \fft are: $t=0.5$\,Myr, $r_c=0.02$\,pc and $\effint=0.01$.  The measured \sfe per \fft is given by the symbol color coding (see right-hand-side palette).  Note that this one is logarithmic and covers three orders of magnitude, from $\effmeasdt = 0.01$ up to $\effmeasdt = 10$.  The lower limit thus corresponds to the intrinsic \sfe per free-fall time actually used in the simulations.  As discussed in Paper~I, the measured \sfe per \fft gets higher if, all other parameters being kept the same, the initial density profile gets steeper (higher $p_0$), or if the $r_c/r_{clump}$ ratio gets smaller, or if the clump mass gets lower (hence a lower volume density and a slower decrease with time of the magnification factor).   Plain circles (triangles) depict models which have converted less (more) than half of their initial gas mass into stars by the elapsed time-span $t=0.5$\,Myr. 

The right panels present, for the same parameter space, the corrected \sfe per free-fall time $\effcor$, that is, the estimate of the intrinsic \sfe per \fft recovered through the method illustrated in Figure~\ref{fig:corgrid}.  Given a sequence of intrinsic efficiencies ($\effint^{TH}=0.005$ to $0.10$ in steps of $0.005$), the corresponding grid of top-hat models is built and the one model minimizing the vertical distance between its \stf relation and that of the centrally-concentrated clump yields the best estimate $\effcor$.  The vertical distance is measured at the mean gas surface density of the top-hat model.  For the sake of comparison, the same color coding is used in the left and right panels.  In contrast to the left panels, the color of the symbols in the right panels is almost uniformly deep-blue or purple, indicating that the method has successfully recovered the intrinsic \sfe per free-fall time, that is, $\effcor \simeq \effint = 0.01$.  This holds regardless of the initial steepness of the density profile, of the clump mass, or of the clump radius.   This also holds independently of how advanced the \stf process is since the right panels show $\effcor \simeq \effint$ for both the triangles ($SFE_{gl} > 0.50$) and the circles ($SFE_{gl} < 0.50$).       

Figure~\ref{fig:cormap2} presents the equally-good results for clumps with the initial gas central density imposed.  As reminded earlier in this section, such clumps often present less extreme magnification factors and convert therefore over the same time-span $t$ a smaller fraction of their gas mass into stars (that is, triangles are less numerous in Figure~\ref{fig:cormap2} than in Figure~\ref{fig:cormap}).  

\begin{figure}
\begin{center}
\epsscale{1.0}  \plotone{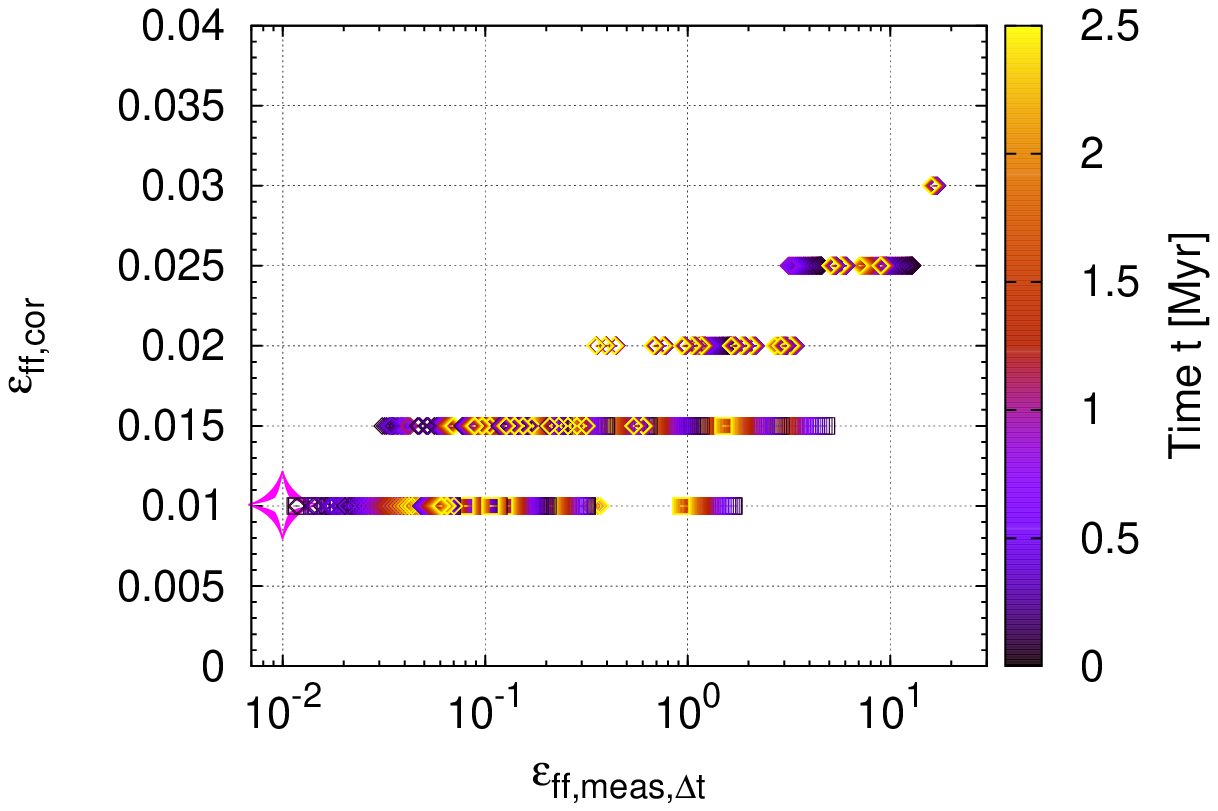}
\caption{Comparison between the measured \sfe per \fft $\effmeasdt$ and our estimate $\effcor$ of its intrinsic counterpart for all tested models.  The time $t$ of the model is color-coded by the right-hand-side palette.  Note the logarithmic scale of the $x$-axis and the linear scale of the $y$-axis.  The magenta four-branch star indicates the intrinsic efficiency actually used in the simulations, $\effint=0.01$}
\label{fig:comp}
\end{center} 
\end{figure}

The total number of tested models amounts to 8000, corresponding to 8 clump masses, 5 radii, 50 \stf time-spans, 2 initial steepnesses $p_0$ of the density profile, and 2 different constraints for the initial gas density profile (either imposed central core radius, or imposed central density).  The corrected efficiency differs from the intrinsic efficiency actually used in the simulations by at most a factor of three (i.e.  $\effint = 0.01\leq \effcor \leq 0.03$: see Figure~\ref{fig:comp}).  This constitutes a great improvement over the globally-measured \sfe per \fft shown in the left panels of Figures ~\ref{fig:cormap} and \ref{fig:cormap2}, where some of these are higher than unity.  

With an estimate of $\effint$ secured, the impact of the initial gas density profile on the clump \stf history can be recovered as the ratio 
\begin{equation}
\zeta_{estimate} = \frac{\effmeasdt}{\effcor}\,,
\label{eq:zetaest}
\end{equation}
with $\effmeasdt$ the "traditional" estimate of the \sfe per free-fall time.  

\section{How to Implement the Method}\label{sec:recip}
In this section, we provide the equations needed to build the "ladder" of \stf relations for top-hat profiles, such as those shown in Figure~\ref{fig:corgrid}.  We then detail how to implement the method in a step-by-step way.
\subsection{Local star-formation relations of top-hat profiles}
For homogeneous models, \stf proceeds at the same pace everywhere (i.e. $\tff(r)$ is independent of the distance $r$ from the clump center).  Therefore, the gas retains a top-hat volume density profile (Figure~\ref{fig:rhogs}), and the forming star cluster has a top-hat volume density profile too.  To build the projected local \stf relation, we need the gas and star {\it surface} density profiles.  When both $\rho_{gas}(r)$ and $\rho_{stars}(r)$ are independent of $r$, the surface density profiles are given by
\begin{multline}
 \Sigma_{gas}(r,dr) \\
= \frac{[r_{clump}^2-(r-dr)^2]^{3/2} - [r_{clump}^2-r^2]^{3/2}}{r_{clump} \cdot dr \cdot (2r-dr)}  \frac{m_{gas}^{TH}}{\pi r_{clump}^2}\,,
\label{eq:sigp0gas}
\end{multline}
and
\begin{multline}
 \Sigma_{stars}(r,dr) \\
 = \frac{[r_{clump}^2-(r-dr)^2]^{3/2} - [r_{clump}^2-r^2]^{3/2}}{r_{clump} \cdot dr \cdot (2r-dr)} \frac{m_{stars}^{TH}}{\pi r_{clump}^2}\;.
\label{eq:sigp0sta}
\end{multline}
In these equations, $\Sigma_{gas}$ and $\Sigma_{stars}$ are the gas and star surface densities measured inside an annulus of inner and outer radii $r-{\rm dr}$ and $r$.  $m_{gas}^{TH}$ and $m_{stars}^{TH}$ are, respectively, the gas and stellar masses of the top-hat model.  We recall that in the method illustrated in Figure~\ref{fig:corgrid}, top-hat models and the centrally-concentrated clump for which an estimate of $\effint$ is sought have the same (mean) volume density.  We therefore assign them the same radius and total mass (i.e. $r_{clump}=r_{clump}^{TH}$ and $m_{clump}=m_{clump}^{TH}$), which implies that: $m_{gas}^{TH}+m_{stars}^{TH} = m_{clump}^{TH} = m_{clump} = m_{gas}+m_{stars}$.

Equations~\ref{eq:sigp0gas} and \ref{eq:sigp0sta} show that three parameters are needed to build the local \stf relation: $r_{clump}$, $m_{gas}^{TH}$ and $m_{stars}^{TH}$.  While observing the centrally-concentrated clump can  provide estimates of $r_{clump}$ and $m_{clump}$, the gas and stellar masses at time $t$ of the top-hat model, $m_{gas}^{TH}$ and $m_{stars}^{TH}$, have to be predicted.  That is, given a model clump of radius $r_{clump}$, initial gas mass $m_{clump}$ and density profile steepness $p_0=0$, what are its stellar mass $m_{stars}^{TH}$ and gas mass $m_{gas}^{TH}$ at time $t$ given an intrinsic \sfe per \fft $\effint^{TH}$?  Since the radius of the clump, thus its volume, is known, we simply need the gas and star volume densities.  These are predicted by Equations~19 and 20 of \cite{par13} which, for an initial uniform gas density $\rho_{TH}$, become
\begin{equation}
\begin{split}
\rho_{gas}^{TH}(t) &= \left( \rho_{TH}^{-1/2} + \sqrt{\frac{8G}{3\pi}} \cdot \epsilon_{ffint}^{TH} \cdot t   \right)^{-2} \\
                              &= \rho_{TH} \left( 1 + \frac{1}{2} \frac{\effint^{TH} \cdot t }{\tff(t=0)} \right)^{-2}\,,
\end{split}
\label{eq:rhog_eq19}
\end{equation}
and
\begin{equation}
\rho_{stars}^{TH}(t) =  \rho_{TH} - \rho_{gas}^{TH}(t)\;.
\label{eq:rhos_eq20}
\end{equation}
The initial gas density $\rho_{TH}$ is known since it equates the mean density of the centrally-concentrated clump $\langle\rho_{clump}\rangle$:
\begin{equation}
\rho_{TH} = \frac{m_{clump}}{\frac{4\pi}{3}r_{clump}^3} = \langle\rho_{clump}\rangle\,,
\label{eq:rhothclump}
\end{equation}
and $\tff(t=0)$ is the free-fall time corresponding to $\rho_{TH}$.

The masses $m_{gas}^{TH}$ and $m_{stars}^{TH}$ then follow from Equations~\ref{eq:rhog_eq19} and \ref{eq:rhos_eq20} as:
\begin{equation}
m_{gas}^{TH}(t)  = \rho_{gas}^{TH}(t) \frac{4 \pi}{3} r_{clump}^3 \,,
\label{eq:mg_eq19}
\end{equation}
and
\begin{equation}
m_{stars}^{TH}(t)  = \rho_{stars}^{TH}(t) \frac{4 \pi}{3} r_{clump}^3 \;.
\label{eq:ms_eq20}
\end{equation}

The gas and star projected density profiles of the top-hat model at time $t$ can now be obtained following Equations~\ref{eq:sigp0gas} and \ref{eq:sigp0sta}, and the local star formation relation follows from plotting the star projected density as a function of its gas equivalent.  \\

\subsection{Step-by-step application of the method}
The method to estimate the intrinsic \sfe per \fft of a centrally-concentrated clump unfolds as follow: \\
- Estimate the radius $r_{clump}$ of the centrally-concentrated clump.  Under the assumption that the clump is in dynamical equilibrium, this is also the clump initial radius; \\
- Estimate the total mass $m_{clump} = m_{gas}+m_{stars}$ enclosed within $r_{clump}$.  Under the assumption that the clump is isolated, the total mass is also the initial gas mass;  \\
- Estimate the time $t$ elapsed since \stf onset in the centrally-concentrated clump; \\
- $t$, $r_{clump}$ and $m_{clump}$ are also the parameters adopted for the top-hat models whose local \stf relations now need to be built; each \stf relation corresponds to one tested value of the intrinsic \sfe per \fft $\effint^{TH}$; \\
- For a given $\effint^{TH}$, Equations~\ref{eq:rhog_eq19} and \ref{eq:rhos_eq20} predict the corresponding gas and star volume densities of the top-hat model following a \stf time-span $t$ (the initial gas volume density $\rho_{TH}$ is known from Equation~\ref{eq:rhothclump}); \\
- The star and gas masses hosted by the top-hat model at time $t$ are then given by Equations \ref{eq:mg_eq19} and \ref{eq:ms_eq20}; \\
- The local star formation relation can now be built as a parametric plot of Equations \ref{eq:sigp0gas} and \ref{eq:sigp0sta};   \\
- The above is to be repeated for a sequence of intrinsic \stf efficiencies per \fft $\effint^{TH}$, which results in a grid of local \stf relations like those shown in Figure~\ref{fig:corgrid}. \\
- Comparing the just obtained model grid with the local \stf relation of the observed centrally-concentrated clump yields an estimate of its intrinsic \sfe per free-fall time. \\

\section{Discussion}\label{sec:disc}
\subsection{$t$-Uncertainties propagate as $\effint$-uncertainties}\label{ssec:disc}

Equations 19 and 20 in \citet{par13} show that, for a given gas initial volume density, the evolutionary stage depends on the product of the \sfe per \fft and of time, $\effint t$ (see also Figure~\ref{fig:sfl1}).  That is, moving a \stf episode "forward" can be done either by considering a longer \stf time-span, or by adopting a higher intrinsic \sfe per free-fall time.  Similarly, Equation~\ref{eq:rhog_eq19} shows that the key parameter associated to each \stf relation of the top-hat grid is not $\effint^{TH}$, but the product $\effint^{TH} \cdot t$.   In Figure~\ref{fig:corgrid}, we have inferred that the best solution is always provided by the "rung" corresponding to $\effint^{TH}=0.01$, namely, the rung which is the closest to the \stf relation of the centrally-concentrated clump (see also bottom panel of Figure~\ref{fig:sfl1}).  However, this holds only if the \stf time-span $t$ has been reliably estimated ($t=0.5$\,Myr in Figure~\ref{fig:corgrid}).  If it has been overestimated by, say, a factor of 4 (i.e. $t_{obs} = 2$\,Myr, with $t_{obs}$ the assumed \stf time-span), then the \sfe per \fft is underestimated by a factor of 4 and $\effcor \simeq 0.0025$.  Conversely, underestimating the \stf time-span by, say, a factor of 5 (i.e. $t_{obs} = 0.1$\,Myr is assumed) results in overestimating the \sfe per free-fall time, in this case by a factor of 5 and $\effcor \simeq 0.05$.  In fact, all top-hat models with $\effcor t_{obs} = \effint t = 0.01 \cdot 0.5 = 0.005$\,Myr provide good matches of the local \stf relation of the centrally-condensed clump.  Uncertainties in the assumed \stf duration are therefore directly reflected as uncertainties in the intrinsic \sfe per freefall time.

\subsection{Clouds vs. Clumps}\label{ssec:clcl}
At this stage, it is important to note that the method has been devised for {\it individual} molecular clumps, {\it not} for entire molecular clouds.  Molecular clouds consist of diffuse gas in which several denser molecular clumps are embedded.  The mean volume density may vary from clump to clump, implying that the clumps have different freefall times and evolutionary paces.  They may also have started to form stars at different times.  As a result, the local \stf relation of a molecular cloud is an assembly of local \stf relations corresponding to \sfing sites at different evolutionary stages, yielding thereby different vertical locations in the ($\Sigma_{gas}$, $\Sigma_{stars}$) parameter space.  The local \stf relation of a cloud can thus be severely thickened and "blurred" \citep[see e.g. Figure~6 in][]{par14p}, thereby preventing a proper comparison with a grid of top-hat models.  This effect pops up nicely when comparing the observed local \stf relations of the Orion and Ophiuchus molecular clouds, as obtained by \citet{gut11}.  While the stellar content of the Ophiuchus cloud is dominated by one embedded cluster (their Figure~6), the Orion cloud is a large collection of numerous individual \sfing regions (their Figure~4).  The result for Ophiuchus is a fairly neatly defined local \stf relation, while the \stf relation of the Orion molecular cloud looks like a broad cloud of points, almost two orders of magnitude thick in stellar surface density (their Figure~9).  It is therefore crucial that observers, when collecting local surface densities of gas and stars for molecular clouds, split their data into well-defined \sfing sites corresponding to the smaller-scale of denser molecular clumps.  Only then can the method be meaningfully applied to observational data sets.  Should the mass and size of these clumps be known, it then becomes possible to constrain their evolutionary stage (the $\effint t$ parameter) by matching them to top-hat models such as those described in this paper.      \\

\section{Conclusions}\label{sec:conc}

Molecular clumps have a higher \sfr when they present a volume density gradient than when they are uniform in density \citep{tan06,gir11a,cho11,elm11,par14p,par19}.  This implies that the \sfe per \fft that is measured for such clumps (Equation~\ref{eq:effmeas}) is also higher than what would be measured if their gas was of uniform density.  This {\it measured} \sfe per freefall time $\effmeas$ is a global quantity since it depends on the \stf rate of clumps, and on the mean density (hence freefall time) and mass of their gas.  
The efficiency of the clump top-hat equivalent is defined as the {\it intrinsic} \sfe per freefall time $\effint$.  That is, this is the \sfe per \fft that would be measured if clumps had no gas density gradient.  For a centrally-concentrated clump, the intrinsic efficiency $\effint$ is also the efficiency characterizing the \stf activity of any clump region small enough to be considered of uniform density (i.e. a region that is small enough so that it does not "see" the clump density gradient).  This is for instance the case of the individual spherically-symmetric shells of gas of which a clump is made   \citep[see Equation 4 in][]{par19}.  The ratio between the measured (equivalently global) and intrinsic (equivalently local) \stf efficiencies per \fft defines the so-called {\it magnification factor} $\zeta$.  Its name stems from $\zeta$ being also the ratio between the \sfr of a centrally-concentrated clump and the \sfr of its top-hat equivalent \citep[see Equation~\ref{eq:zetadef} of this paper and ][]{par19}.              
The implications are that, even for a fixed $\effint$, its measured counterpart  $\effmeas$ present wide variations, reflecting the diversity of clump inner structures rather than variations in the \stf process itself.  Intrinsic and measured efficiencies are equal (similar) for top-hat (shallow) gas density profiles only.  
  
That the degree of central concentration of a molecular clump contributes to its measured \sfe per free-fall time $\effmeas$, thereby masking the intrinsic efficiency $\effint$ at work inside its constituent shells, leaves us with an annoying degeneracy.  For instance, when the measured \sfe per freefall time of a clump is high, one cannot {\it a priori} disentangle whether this results from a high intrinsic \sfe per \fft  or from a clump steep density gradient.  In the first case, the gas is highly efficient at forming stars, and would remain so even if the clump gas were of uniform density.  In the second case, this is the clump gas central concentration, embodied by the magnification factor $\zeta$, which drives the high measured \sfe per freefall time of the clump.   

In this paper, we have presented a method allowing one to lift this degeneracy.  It builds on the local \stf relation, which relates the densities in gas and stars at a given radial location inside the \sfing clump (hence the term 'local').  The method requires therefore spatially-resolved observations of molecular clumps.  Global (i.e. clump-averaged) data are not enough.  
The key idea on which the method hinges is that steepening the volume density profile of a clump, starting from a top-hat model, stretches out its local \stf relation while retaining the vertical normalization of the top-hat model (see top panel of Figure~\ref{fig:sfl1}).  
This suggests that to estimate the intrinsic \sfe per \fft $\effint$ of a centrally-concentrated clump, one can compare its  local \stf relation with a grid of top-hat models of identical mass, radius, \stf time-span, but with their own \stf efficiencies per freefall time $\effint^{TH}$.  $\effint$ follows from selecting the top-hat model which minimizes the vertical offset between its \stf relation and that of the centrally-concentrated clump.   
 This property is also valid in the parameter space of gas- and star-surface densities (see bottom panel of Figure~\ref{fig:sfl1}), making the method applicable to observational data sets.  Knowledge of the clump gas initial  density profile is not required.  Only its total mass, radius and \stf time-span are.  With these data, one can build the corresponding sequence of \stf relations for top-hat profiles (see the "ladders" made of star-formation relations in Figure~\ref{fig:corgrid}).

We have systematically applied our method to the model clumps with an initially steep density profile calculated in \citet{par19}.  By "steep", we mean power-law density profiles with a logarithmic slope of $-3$ or steeper initially.  Assuming that the \stf time-span is known, we recover, to better than a factor of 3, the intrinsic \sfe per \fft $\effint$ actually used in the simulations.  For comparison, it should be noted that $\effmeas$ differs from $\effint$ by up to three orders-of-magnitude in the most centrally-concentrated models (see Figure~\ref{fig:comp}).
We have provided a step-by-step description of the method, and the equations that its implementation requires (Section~\ref{sec:recip}).   
Uncertainties in the \stf time-span $t$ are reflected as $\effint$-uncertainties since the clump evolutionary stage depends on their product \citep[see Equation 19 in][]{par13}.  
We stress that the method must be applied to individual molecular clumps, rather than to their host molecular clouds,  given that clouds consist of several clumps, each with its own \stf time-span and initial gas density.  A collection of clumps therefore leads to "piling-up" several \stf relations (Section~\ref{sec:disc}). 

Once an estimate of the intrinsic \sfe per \fft $\effint$ has been secured, it can be combined to the "traditional"/globally-measured \sfe per \fft $\effmeas$ to assess the impact that the gas density gradient of a clump has had on its star-formation history (Equation~\ref{eq:zetaest}).  Combining the method presented in this contribution with the spatial resolution of the Atacama Large Millimetre Array will allow us to investigate whether the intrinsic \sfe per \fft of Galactic molecular clumps varies as a function of environment.



\acknowledgments
G.P. is grateful to Anna Pasquali and Douglas Heggie for stimulating discussions while working on this manuscript. G.P. acknowledges funding by the Deutsche Forschungsgemeinschaft (DFG, German Research Foundation -- Project-ID 138713538 -- SFB 881 ("The Milky Way System", subproject B05).







\end{document}